\documentclass[aip,jap, reprint, twocolumn, showpacs, superscriptaddress]{revtex4-1}
\usepackage[T1]{fontenc}
\usepackage[utf8]{inputenc}
\usepackage{graphicx}
\usepackage{pstricks}
\usepackage{array}
\usepackage{natbib}
\usepackage{amsmath}
\usepackage{pifont}
\usepackage{dsfont}
\usepackage{multirow}
\usepackage{amssymb}
\usepackage{wasysym}
\usepackage[titletoc,toc,title]{appendix}

\def\be{\begin{equation}}
\def\ee{\end{equation}}

\newcommand{\cnst}{Center for Nanoscale Science and Technology, National
Institute of Standards and Technology, Gaithersburg, MD 20899}
\newcommand{\umd}{Maryland NanoCenter, University of Maryland, College
Park, MD 20742}

\begin{document}

\title{Probing surface recombination velocities in semiconductors using two-photon microscopy}

\begin{abstract}
The determination of minority-carrier lifetimes and surface recombination
velocities is essential for the development of semiconductor technologies such
as solar cells. The recent development of two-photon time-resolved
microscopy allows for better measurements of bulk and subsurface
interfaces properties. Here we analyze the diffusion problem
related to this optical technique. Our three-dimensional treatment enables us to
separate lifetime (recombination) from transport effects (diffusion) in the
photoluminescence intensity. It also allows us to consider surface recombination
occurring at a variety of geometries: a single plane (representing
an isolated exposed or buried interface), two parallel planes (representing two
inequivalent interfaces), and a spherical surface (representing the enclosing
surface of a grain boundary).  We provide fully analytical results and scalings
directly amenable to data fitting, and apply those to experimental data
collected on heteroepitaxial CdTe/ZnTe/Si.
\end{abstract}

\author{Benoit Gaury}
\affiliation{\cnst}
\affiliation{\umd}
\author{Paul Haney}
\affiliation{\cnst}

\date{\today}

\maketitle

\section{Introduction}
The minority-carrier lifetime may be considered the
most critical parameter for photovoltaic materials. However, in polycrystalline
materials like CdTe, the exact contribution of
bulk, grain boundaries and other interfaces to recombination losses is still
unclear.  Optical techniques, such as time-resolved photoluminescence (TRPL), have
been developed to probe the bulk lifetime and surface recombination
velocities of direct bandgap materials.  These experiments consist of optically
generating electron-hole pairs and observing the time-dependence of the photons
emitted from radiative recombination. The time constant of the
signal decay contains information about bulk and surface recombination.

\begin{figure}
    \includegraphics[width=0.48\textwidth]{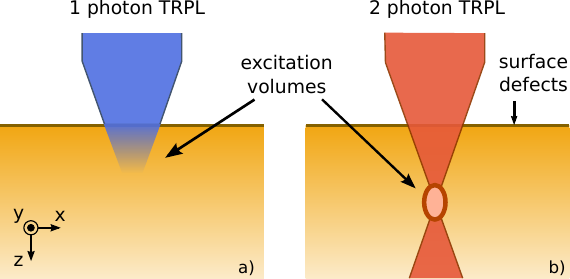}
    \caption{\label{fig:2P} Schematics of the one- and two-photon excitation
    microscopy techniques. (a) The incident beam is absorbed exponentially away
    from the surface. (b) Charge carriers are excited at the focal point of the
    incident beam, under the surface of the sample.} 
\end{figure}

TRPL measurements are most commonly conducted using 
one-photon excitation.  In this case the energy of the incident photons is
larger than the semiconductor bandgap so that a single photon generates an
electron-hole pair (or exciton). The absorption in the bulk decays
exponentially away from the sample surface, as shown in Fig.~\ref{fig:2P}a).  More recently, two-photon TRPL has
been developed and applied to photovoltaic materials~\cite{Wang_1998,
Zhong_2006, Ma_2013}.  In this case,  the energy
of the incident photons is smaller than the bandgap such that multiple photon
absorptions are required to excite an electron-hole pair. This non-linear
absorption process is obtained by focusing a laser beam under the sample
surface, as shown in Fig.~\ref{fig:2P}b).  The procedure allows one to generate
carriers at any desired depth inside the material, so that surface
and bulk contributions can be disentangled in the TRPL response. We refer to the
literature~\cite{Barnard_2013,Kuciauskas2014}  for details on the operating
principle and experimental setups.

While there exists an abundant literature on the modeling of one-photon TRPL
measurements~\cite{Boulou_1977, Hooft_1986, Ahrenkiel_1989}, theoretical works
for two-photon TRPL are still scarce and limited to numerical
studies~\cite{Kanevce_2015}. The present work builds on the extensive body of
mathematical analysis developed from earlier investigations of
one-photon TRPL, which assumed an optical excitation that decays away from the
sample surface into the bulk according to Beer’s law.  

In this work we analyze the diffusion problem related to the
two-photon TRPL microscopy technique. Because PL intensities often depend
on material parameters in non-trivial ways, we use the analytical 
solutions of our modeling to propose scalings and experimental procedures which enable
convenient extraction of these parameters. We are considering a model of excess
carrier diffusion in three dimensions in the low-injection regime (i.e. with
first order recombination), with recombination that is first order in
carrier density. We provide general results for the minority-carrier
concentration and the photoluminescence (PL) intensity for an excitation of
arbitrary spatial dependence.  Because the free carriers can diffuse away from
the photon collection region before they radiatively recombine, the PL signal is
strongly affected by the dimensions of generation and collection volumes. This
makes our 3D treatment relevant to identifying the impact of the free carriers
transport on the time decay of the PL intensity. The paper is organized as
follows: We present our model for the transport of optically-generated minority
carriers in Sec.~\ref{model}. The Green's function of the diffusion problem is
introduced, and we show how it relates to the minority-carrier density.
Section~\ref{semi-infinite} presents the case of a volume bounded by a single
planar boundary with enhanced recombination.  We provide the solution of the 3D
diffusion equation for an arbitrary generation and collection volume.  We then
consider the special case of point-like excitation and collection, and use this
result to derive expressions which can be conveniently used for data fitting.
We end the section using this solution to fit the experimental data obtained on
heteroepitaxial CdTe/ZnTe/Si in Ref.~\onlinecite{Kuciauskas2014}. A second
planar boundary in the axial direction is added in Sec.~\ref{2planes}. Finally,
we investigate the spherical diffusion of minority carriers in Sec.~\ref{sphere}
to determine the surface recombination velocity of an enclosing grain boundary
surface.

\section{Model for the laser beam induced minority-carrier transport}
\label{model}

Without loss of generality we develop our analysis for a p-type material. We
model the transport of optically-induced electrons with the
time-dependent diffusion equation 
\be
    \frac{\partial n}{\partial t}(\mathbf{r},t)- D\Delta n(\mathbf{r},t) +
    \frac{n(\mathbf{r},t)}{\tau} = g(\mathbf{r})\delta(t),
    \label{3Dproblem}
\ee
where $n(\mathbf{r}, t)$ is the electron concentration, 
$D$ is their diffusion constant, $\tau$ their bulk lifetime
and $g(\mathbf{r})\delta(t)$ is the carrier generation, assumed
instantaneous. The recombination rate is taken to first
order in minority-carrier density, so that Eq.~(\ref{3Dproblem}) is valid only
in the low-injection regime. For the purpose of calculating the
distribution of electrons after a laser pulse, we introduce the Green's function
$G(\mathbf{r}, \mathbf{r'}, t)$ that satisfies
\be
    \frac{\partial G}{\partial t}(\mathbf{r}, \mathbf{r'},t)- D\Delta G(\mathbf{r}, \mathbf{r'},t) +
    \frac{G(\mathbf{r}, \mathbf{r'},t)}{\tau} = \delta(\mathbf{r}-\mathbf{r'})
    \delta(t).
\label{GF}
\ee
Upon solving Eq.~(\ref{GF}) with the appropriate boundary conditions, the
electron density is obtained with the convolution
\be
    n(\mathbf{r}, t) = \int \mathrm{d}\mathbf{r'} \ G(\mathbf{r},
    \mathbf{r'}, t) g(\mathbf{r'}),
\label{convol}
\ee
and integrating the above density over a collection volume $V$ yields the PL intensity
\be
    I(t) = \frac{1}{\tau_r} \int_V \mathrm{d}\mathbf{r} \ n(\mathbf{r},t),
    \label{PL_signal}
\ee
where $\tau_r=1/Bp_0$ is the radiative lifetime ($B$: radiative recombination
coefficient of the material, $p_0$: hole doping).  The lateral collection area
is set by the spot size, while the generation volume is contained within the
spot size and is generally smaller (the generation volume is determined in part
by the laser fluence~\cite{Shao_2011}). The volumes in Eqs.~(\ref{convol})
and~(\ref{PL_signal}) are therefore different. For simplicity, we take these
volumes to be equal in the experimental procedures presented in
Secs.~\ref{semi-infinite} and~\ref{sphere}.  The results of the successive
integrations of the Green's function (one for the carrier density and a second
one for the PL intensity) become quickly very messy in the rare cases where we
can compute them analytically. However, this is easily done numerically. 

The model introduced here has several limitations. First, it does not include
the self-absorption of radiatively emitted photons with the subsequent
electron-hole pair creation, or so-called photon recycling~\cite{Stern_1974}.
Photon recycling can alter the measured PL lifetime only when the radiative
recombination mechanism is not negligible.  For direct band gap polycrystalline
semiconductors with low majority-carrier densities
($10^{15}-10^{16}~\mathrm{cm^{-3}}$), the Schockley-Read-Hall recombination
mechanism dominates. Photon recycling can therefore be neglected in these
materials. Second, the model leaves out space charge effects caused by
surface-induced electric fields and differences in electron and hole mobilities.
These non-linear effects have been studied numerically~\cite{Kanevce_2015}.
Finally, while refraction and diffraction are not limitations of this model per
se, we will discuss how the optics influences the extraction of physical
parameters from experimental data in the next section.

We will now proceed with the resolution of Eq.~(\ref{GF}) for various boundary
conditions.

\section{Two-photon TRPL in a semi-infinite system}
\label{semi-infinite}
We discuss the case of a volume bounded by a single planar surface with enhanced
recombination.  This surface can be the sample surface, as depicted in
Fig.~\ref{fig:2P}, or any buried subsurface or material interface. We provide
the analytic solution to the 3D diffusion problem. Next we introduce the ratio
of the PL signals for excitations deep in the bulk and near the surface, and
show that it varies linearly with time.  The slope of this linear variation can
be used to determine the diffusion constant and the surface
recombination velocity.

\bigskip
Eq.~(\ref{GF}) is solved in the half volume $z \geqslant 0$ with the
boundary conditions determined by the surface recombination velocity $S$ at
$z=0$,
\begin{align}
    \label{BC1}
    D\frac{\partial n}{\partial z} = S n, &\ \ z=0\\
    n = 0, &\ \ z\rightarrow+\infty
    \label{BC2}
\end{align}
The solution reads
\begin{align}
    G&(x,x',y,y',z,z',t) = \frac{e^{-t/\tau}}{2}  \frac{e^{-\frac{(x-x')^2}{4
    Dt}}}{2\sqrt{\pi Dt}} \frac{e^{-\frac{(y-y')^2}{4
    Dt}}}{2\sqrt{\pi Dt}}
    \nonumber\\
    &\Bigg[ \frac{e^{-\frac{(z-z')^2}{4Dt}} +
    e^{-\frac{(z+z')^2}{4Dt}}}{\sqrt{\pi Dt}} \nonumber\\
    & - 2\frac{S}{D}
    e^{{\frac{S}{D}(z+z')}+\frac{S^2}{D}t} \mathrm{erfc}\left(\frac{z+z'}{2\sqrt{Dt}} +
    S\sqrt{\frac{t}{D}}  \right) \Bigg],
    \label{G_final}
\end{align}
where $\mathrm{erfc}$ is the complementary error function. A derivation can be
found in Ref.~\onlinecite{Roosbroeck_1955} and in Appendix~\ref{appendix1}.
Direct numerical integration of Eq.~(\ref{convol}) with the Green's function
Eq.~(\ref{G_final}) over the generation/collection region yields the
time-dependent PL intensity for this geometry.  Eq.~(\ref{G_final}) is an exact
solution to the problem and, with the aforementioned numerical work, can be used
for data fitting. However the expression is complicated and not especially
intuitive.

To gain insight into the role of diffusion and surface recombination, we
consider a limiting case of Eq.~(\ref{G_final}), for which the
generation/collection regions are both point sources.  These are strong
approximations but they give good insights for i) the shape of the PL signal, in
particular the behavior at short times, and ii) possible procedures to determine
the recombination velocity $S$.  The point source is positioned at $(0,0,z_0)$
with amplitude $g_0$
\be
    g(x,y,z) = g_0 \delta(x)\delta(y)\delta(z-z_0).
\ee
Applying Eq.~(\ref{convol}) yields the spatial distribution of electrons in the system 
\be
    n(x,y,z,t) = G(x,0, y,0,z,z_0,t).
    \label{nn}
\ee
The PL intensity of photons originating from $z=z_0$ is given by
\begin{align}
    I(t) = &\frac{g_0}{\tau_r}
    \frac{e^{-t/\tau}}{8\pi Dt}
    \Bigg[ \frac{1 +
    e^{-z_0^2/(Dt)}}{\sqrt{\pi Dt}} \nonumber\\
    & - 2\frac{S}{D}
    e^{{\frac{S}{D}2z_0}+\frac{S^2}{D}t} \mathrm{erfc}\left(\frac{z_0}{\sqrt{Dt}} +
    S\sqrt{\frac{t}{D}}  \right) \Bigg].
    \label{I}
\end{align}
We first consider an excitation in the bulk, far from the surface ($z_0\gg L_d$,
$L_d=\sqrt{D\tau}$: diffusion length).
In this case we can ignore the second term in square brackets in Eq.~(\ref{I}),
and get
\be
    I_b(t) = \frac{g_0}{\tau_r} \frac{e^{-t/\tau}}{8(\pi Dt)^{3/2}}.
    \label{bulk}
\ee
\begin{figure}[b]
    \includegraphics[width=0.48\textwidth]{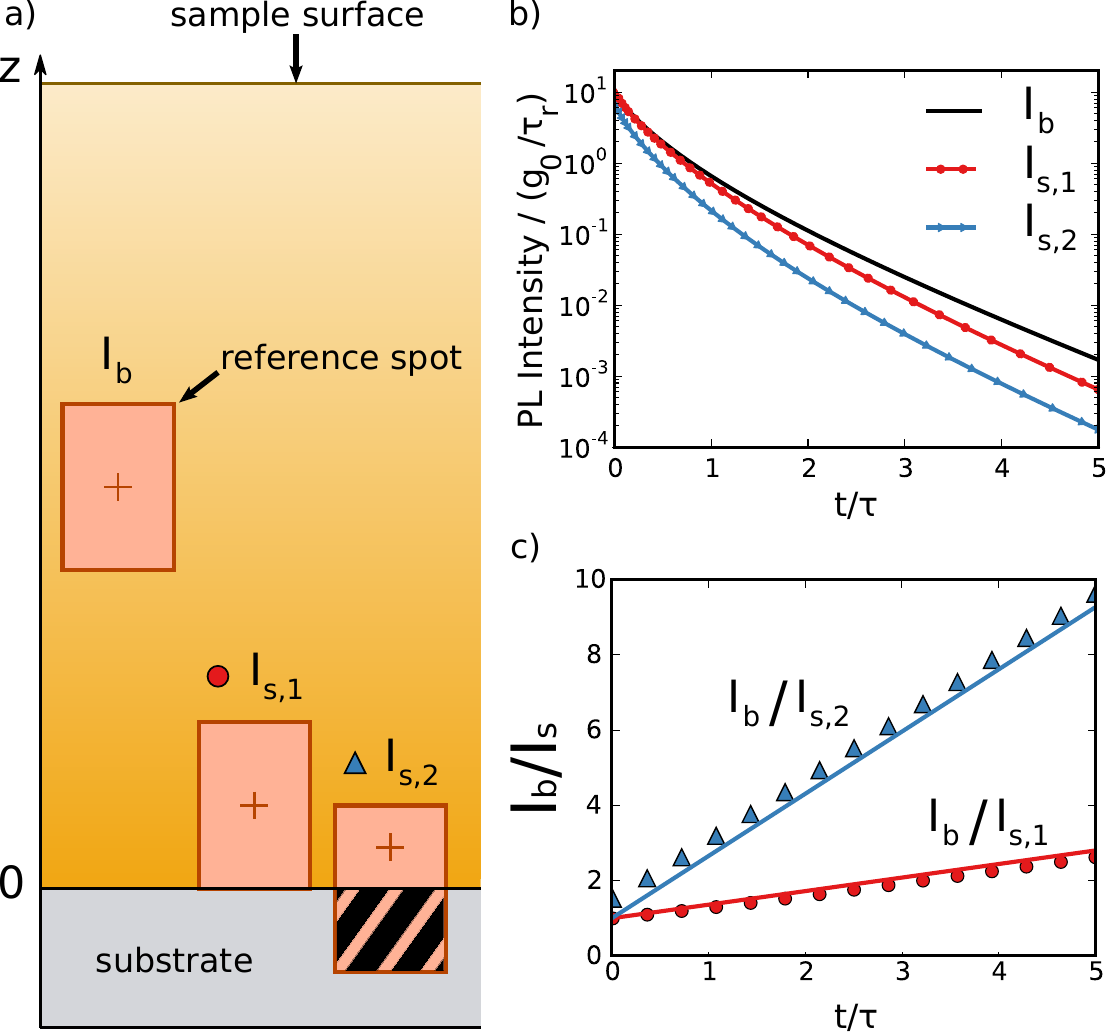}
    \caption{\label{1plane} Procedure to determine the surface recombination
    velocity and the diffusion constant at a sample substrate/active layer
    interface.  (a) Schematic of the sample with three focal points for the
    generation/collection region respectively in the
    bulk and at the substrate/active layer interface. (b) PL intensities as a
    function of time corresponding to the three excitations obtained for uniform
    rectangular shaped generation/collection volumes, as shown in a). The curves
    are numerical integrations of Eq.~(\ref{convol}). The reference spot in the
    bulk (black, $I_b$) has lateral size $2L_d$ and axial size $3L_d$.
   The axial sizes of the spots in the
    active layer at the
    substrate/system interface are $3L_d$ (red with dots, $I_{s,1}$) and $2L_d$
    (blue with triangles, $I_{s,2}$).
    (c) Ratio of the bulk and interface PL intensities. Symbols correspond to
    the ratio of the data plotted in b). The continuous lines correspond to
    Eq.~(\ref{scaling}) with $z_0$ indicated by the red cross inside each spot
    in a).  $z_0$ values are $L_d$ (blue, triangles) and $1.5L_d$ (red, dots).
    The reference spot has $z_0=9L_d$. The surface recombination velocity was
    chosen such that $S=6v$ ($v$: diffusion velocity).}
\end{figure}
This result indicates that the diffusion of carriers mostly influences the PL
intensity at short times ($e^{-t/\tau}\approx 1$), giving an {\it algebraic}
form to the decay instead of an exponential one. At long times, the bulk
lifetime dominates the PL signal but the PL decay is still not purely
exponential.  One recovers a purely exponential decay when the 
laser spot size is much greater than the diffusion length. 
Formally this amounts to integrating Eq.~(\ref{G_final})
over all spatial directions ignoring the second term in the square brackets.
Next we consider an excitation near the surface ($z_0 \ll St$, $z_0 \ll
\sqrt{Dt}$), and focus on long times ($t>\tau$). In this case, we obtain 
\be
    I_s(t) = \frac{g_0}{\tau_r}\frac{(z_0+D/S)^2 e^{-t/\tau}}{8 \pi^{3/2} (Dt)^{5/2}}.
    \label{surf}
\ee
Although Eqs.~(\ref{bulk}) and (\ref{surf}) were derived for a point source
generation/collection region, we find that for a finite generation/collection
region of volume $V$, the same relations hold with an additional prefactor of
$V^2$. This assumes a uniform carrier generation, and times long enough so
that carriers have diffused outside the collection volume, that is $\sqrt{Dt}\gg V^{1/3}$.
We notice that the time dependence of the signal for near-surface excitations
$I_s(t)$ has an additional factor of $1/t$ compared to the bulk excitation signal
$I_b(t)$.  For this reason we propose that forming the ratio of these PL signals
may be a convenient way of estimating material parameters.  Including the
generation/collection volume factors we discussed above, the ratio is given by
\be
    I_b(t) / I_s(t) = 1 + \left(\frac{V_b}{V_s}\right)^2
    \frac{Dt}{(z_0+D/S)^2},
    \label{scaling}
\ee
where $V_b$ and $V_s$ are the generation/collection
volumes in the bulk and close to the surface respectively; $z_0$ is now the
distance of the center of the uniform generation volume from the surface.

To test the accuracy and demonstrate the use of Eq.~(\ref{scaling}), we perform numerical
calculations in which we compare Eq.~(\ref{scaling}) to the full numerical
integration of Eqs.~(\ref{convol}) and (\ref{G_final}).  Fig.~\ref{1plane}(a)
shows the geometry; we choose three focal points for the generation/collection
region: one in the bulk, and two near a buried sample-substrate interface.  Note
that as the generation/collection region approaches the interface, a portion of
it falls outside the sample and does not contribute to the generation or
collection (hatching in the right spot of Fig.~\ref{1plane}(a)).
Fig.~\ref{1plane}(b) shows the PL time traces obtained by direct numerical
integration.  As expected, the signal decreases more quickly for excitations
near the sample-substrate interface ($I_{s,1}$ and $I_{s,2}$).  The symbols of
Fig.~\ref{1plane}(c) shows the ratios between the two subsurface excitations and
the bulk excitation.  The solid lines of Fig.~\ref{1plane}(c) show
Eq.~(\ref{scaling}), demonstrating good agreement with the full numerical
calculations.  Given a similar set of experimental data (one bulk and two
near-surface excitations, or two values of $z_0$), one can form the ratio of the
PL signals of excitations in the bulk and near the surface, and use the
resulting slopes to estimate $S$ and $D$. 

\bigskip
We now apply the formulas derived above to fit the experimental data taken from
Ref.~\onlinecite{Kuciauskas2014}. 
\begin{figure}[t]
\includegraphics[width=0.48\textwidth]{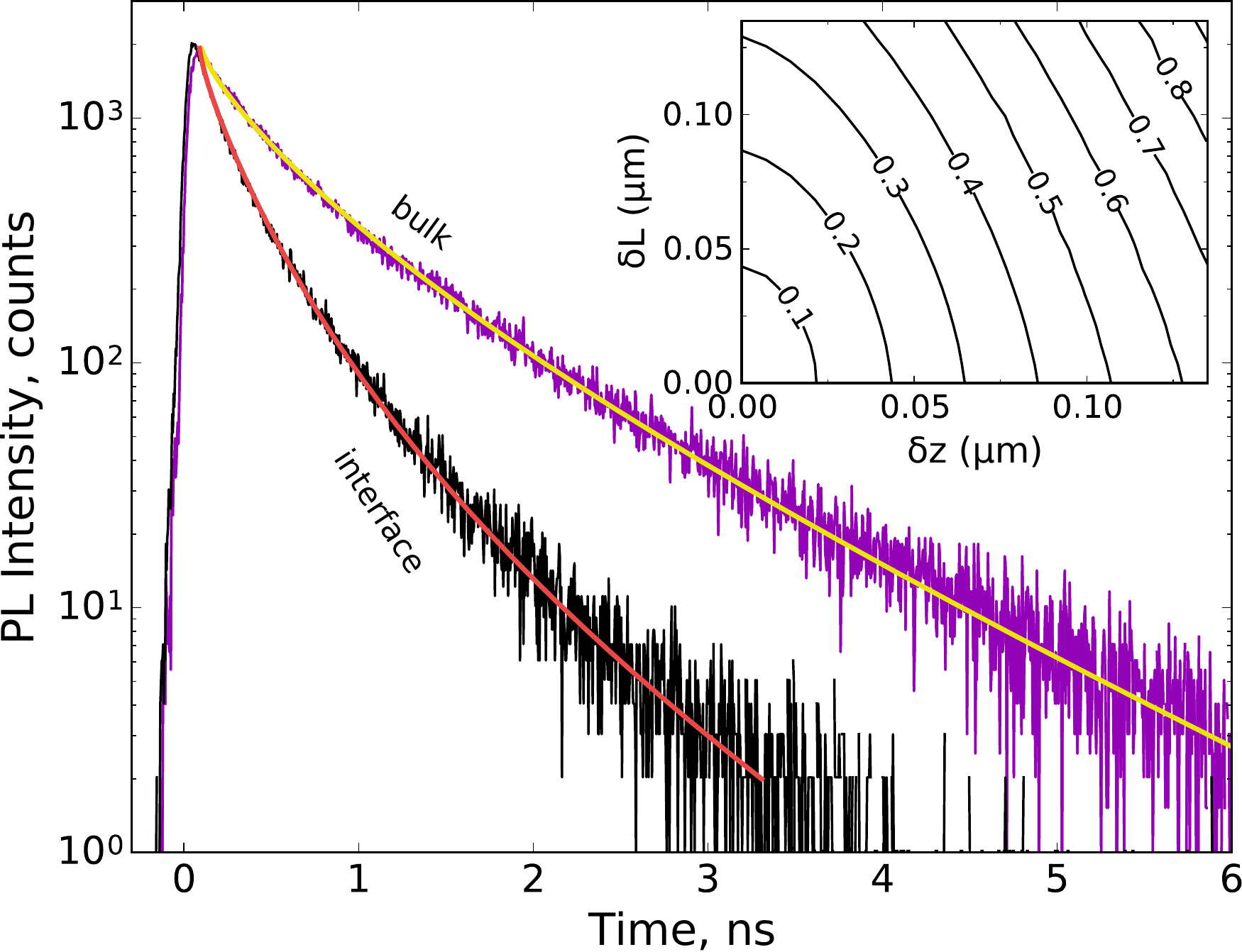}
\caption{\label{fitting}Fitting of the experimental data of Fig.~6 from
Ref.~\onlinecite{Kuciauskas2014}. The purple and black traces are two-photon
TRPL measurements taken respectively in the bulk and close to the
heteroepitaxial CdTe/ZnTe/Si interface of a CdTe layer grown on Si substrate. The yellow and red curves correspond to
numerical evaluations of Eq.~(\ref{convol}) with the Green's
function Eq.~(\ref{G_final}). The generation and collection volumes are the same
with lateral size (x and y) $2~\mathrm{\mu m}$,
and axial sizes $5.5~\mathrm{\mu m}$ (yellow) and $1.2~\mathrm{\mu m}$ (red). The
center of the spot is at $8.75~\mathrm{\mu m}$ from the interface for the yellow
curve and $0.6~\mathrm{\mu m}$ for the red curve. Parameters obtained after
least square fitting:
$S=4.4\times10^5~\mathrm{cm~s^{-1}}, D=6.5~\mathrm{cm^2~s^{-1}},
\tau=1.6~\mathrm{ns}$. Inset: Relative uncertainty $\delta S /S$ on the
extracted values
of $S$ as a function of the error on the position of the center of the spot
($\delta z$), and its axial length ($\delta L$).}
\end{figure}
Kuciauskas and coworkers reported on one- and
two-photon TRPL studies of CdTe grown on Si(211) substrates. Here we focus on
the characterization of the buried heteroepitaxial CdTe/ZnTe/Si interface
presented in Fig.~6 of the article (CdTe layer thickness: $17.5~\mathrm{\mu m}$). In
particular we consider two spot positions: in the bulk (purple) and close to the
interface (black).
Fig.~\ref{fitting} shows the experimental data with our fitting curves.  Our
theoretical model for the generation volume consists of a uniform generation of
rectangular shape. The generation and collection volumes were kept the same with
lateral length $2~\mathrm{\mu m}$.  The axial spot size was $5.5~\mathrm{\mu m}$
in the bulk (yellow) and $1.2~\mathrm{\mu m}$ at the interface (red).  We
attribute this reduction to a large amount of the laser spot being partially focused on
the silicon substrate, hence reducing significantly the number of electrons
generated in the CdTe layer.  The material parameters obtained after a least
square fitting are $S=4.4\times10^5~\mathrm{cm~s^{-1}}$,
$D=6.5~\mathrm{cm^2~s^{-1}}$ and $\tau=1.6~\mathrm{ns}$. This gives the
diffusion length 
$L_d=1.0~\mathrm{\mu m}$. These values are to be compared to the ones found in
Ref.~\onlinecite{Kuciauskas2014}: $S=(6\pm 2)\times 10^5~\mathrm{cm~s^{-1}}$, $D=17.0\pm
0.3~\mathrm{cm^2~s^{-1}}$.   

There is some degree of insensitivity of the fitting parameters. For example, we
find that good fitting persists for variations of the diffusion constant, the
bulk lifetime and the surface recombination by 13~\%, 7.5~\% and 30~\%
respectively. Furthermore, the inset of Fig.~\ref{fitting} shows the uncertainty
on the extracted surface recombination as the uncertainties of the center of the
laser spot ($\delta z$) and its axial length ($\delta L$) vary. An error on both
quantities of only $0.1~\mathrm{\mu m}$ results in a $60~\%$ relative error on
$S$. This sensitivity implies that a precise knowledge of the optical generation
and collection volumes is required for a precise extraction of physical
parameters from experimental data. 

\begin{figure}[t]
\includegraphics[width=0.48\textwidth]{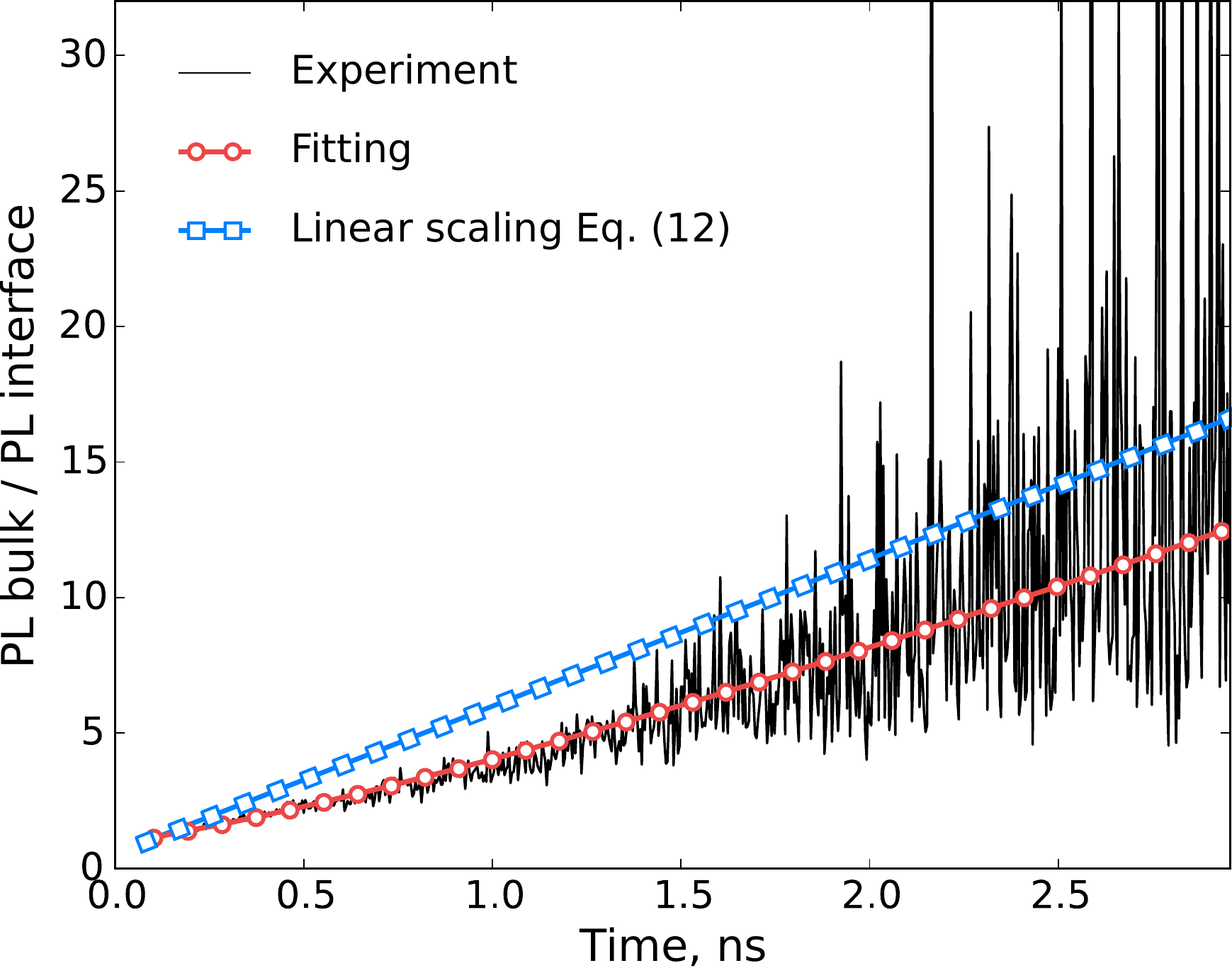}
\caption{\label{fig:scaling} Ratio of bulk and interface photoluminescence
intensities as a function of time. The black (red-white dotted) continuous line corresponds
to the ratio of the experimental (numerical) data presented in
Fig.~\ref{fitting}. The blue line with square symbols corresponds to Eq.~(\ref{scaling})
with the parameters used in Fig.~\ref{fitting}.}
\end{figure}
Fig.~\ref{fig:scaling} compares the scaling proposed in Eq.~(\ref{scaling}) to
the ratio of the PL signals taken in the bulk and at the interface.
The time interval ($<3~\mathrm{ns}$) is small compared with the lifetime
$\tau=1.6~\mathrm{ns}$, and Eq.~(\ref{scaling}) applies for $t>\tau$.
Nevertheless, the scaling agrees with the trend of the experimental data. In
addition to the aforementioned  experimental challenges related to optics,
another difficulty is to obtain data at the surface with little noise on a time
scale at least several times the minority-carrier lifetime. This can be
difficult because of the fast decay of the PL intensity caused by the enhanced
recombination occurring near the surface.

\section{Diffusion bounded by two parallel planes}
\label{2planes}
\begin{figure}[h]
\includegraphics[width=0.28\textwidth]{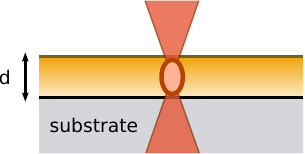}
\caption{\label{2plane_schematic} Schematic of a sample with thickness $d$ on
the same order of the laser spot size.}
\end{figure}

We now turn to a system with finite thickness $d$ in the
z-direction. This situation is especially applicable to thin films of thickness on the
order of the laser spot axial dimension (typically a few $\mathrm{\mu m})$, as
depicted in Fig.~\ref{2plane_schematic}. 
The PL signal is then sensitive to the sample surface as well as the
sample/substrate interface.  We provide the analytic solution to the 3D
diffusion problem studied in the previous section with the additional boundary
condition
\be
    D\frac{\partial n}{\partial z} = -S'n, \ \ z=d
    \label{BC3}
\ee
The solution of Eq.~(\ref{GF}) with the
boundary conditions Eq.~(\ref{BC1}) and Eq.~(\ref{BC3}) is given by
\be
    G(x,x',y,y',z,z',t) =
    e^{-t/\tau}\frac{e^{-\frac{(x-x')^2}{4Dt}}}{2\sqrt{\pi Dt}} 
               \frac{e^{-\frac{(y-y')^2}{4Dt}}}{2\sqrt{\pi Dt}}
    U(z,z',t)
    \label{G_final2}
\ee
with
\be
    U(z,z',t) = \sum_{n>0} A_n(z') A_n(z)e^{-\alpha_n^2Dt},
    \label{series}
\ee
where the coefficients $\alpha_n$ satisfy the transcendental relation
\be
    \tan(\alpha_n d) = \frac{\alpha_n (S+S')D}{\alpha_n^2D^2 - SS'}.
    \label{trans}
\ee
The details of the calculations as well as the expression of the coefficient
$A_n(z)$ are given in Appendix~\ref{appendix2}.  
As with the single planar boundary, the diffusion in
free space results in {\it non-exponential} terms to the PL signal given by the $x$ and
$y$ components in Eq.~(\ref{G_final2}). The transcendental equation
Eq.~(\ref{trans}) has already been found and studied in the context of excess
minority-carrier excitation decay~\cite{Ahrenkiel_1993, Sproul_1994}. The literature
provides approximations for the characteristic time $\tau_1$ corresponding to
the first non-zero root of Eq.~(\ref{trans}). For any values of $S$ and $S'$,
$\tau_1$ can be approximated by~\cite{Scajev_2010}
\be
    \tau_1 = \frac{(\tau_D+\tau_S+\tau_{S'})(\tau_D+\tau_S)(\tau_D+\tau_{S'})}
                  {\left(\tau_D + (\tau_S+\tau_{S'})/2\right)^2},
    \label{approx_tau1}
\ee
with $\tau_D=d^2/\pi^2D$, $\tau_S=d/4S$ and $\tau_{S'}=d/4S'$.
Equation~(\ref{approx_tau1}) encompasses approximations found in earlier works, in
particular in the case where $S=S'$, $\tau_1$ can be approximated by~\cite{Boulou_1977}
\be
    \tau_1 \approx \frac{d^2/D}{\pi^2} + \frac{d}{2S},
\ee
and in the case where $S'=0$ (or equivalently $S=0$) it has been found that~\cite{Sproul_1994}
\be
    \tau_1 \approx \frac{4d^2/D}{\pi^2} + \frac{d}{S}.
    \label{emp1}
\ee
For times long compared to $d^2/(\pi^2D)$ the Fourier series Eq.~(\ref{series})
can be reduced to its first term so that the exponential component of the PL
signal decays with the effective characteristic time
\be
    \frac{1}{\tau_{PL}} = \frac{1}{\tau} + \frac{1}{\tau_1}.
    \label{tau_PL}
\ee
This remains valid whatever the shape of the generation volume (because the
successive integrations on Eq.~(\ref{G_final2}) will not affect the
time-dependent term), as long as the
laser pulse duration is much smaller than all other characteristic times. 

Equation~(\ref{tau_PL}) together with Eq.~(\ref{trans}) (or with the
approximation Eq.~(\ref{approx_tau1})) require prior knowledge of the bulk
lifetime and the diffusion constant in order to access $S$ and $S'$. Assuming
that this condition is met, one would make two devices with different
thicknesses to determine both surface recombination velocities. To simplify the
extraction of the parameters from experimental data, one may consider a laser
beam with lateral dimensions much larger than the diffusion length. This
procedure allows one to formally integrate Eq.~(\ref{G_final2}) over $x$ and
$y$, so that the PL decay becomes purely exponential.  
As the thickness $d$ increases, $\tau_1$ asymptotes to $d^2/\pi^2 D$, and no longer
contains information about the surface; therefore $d$ must be sufficiently small
to extract $S$. Also, $d$ must be chosen so that $1/\tau_1$ is not negligible compared
to $1/\tau$. As an example, with $S=S'=5\times
10^{5}~\mathrm{cm~s^{-1}}$, $D=10~\mathrm{cm~s^{-1}}$ and $\tau=1~\mathrm{ns}$,
one finds that $d\leqslant 2.7~\mathrm{\mu m}$ is needed to obtain $\tau_1
\leqslant \tau$.  

\section{Determination of the surface recombination velocity of a grain boundary}
\label{sphere}
\begin{figure}[t]
    \includegraphics[width=0.35\textwidth]{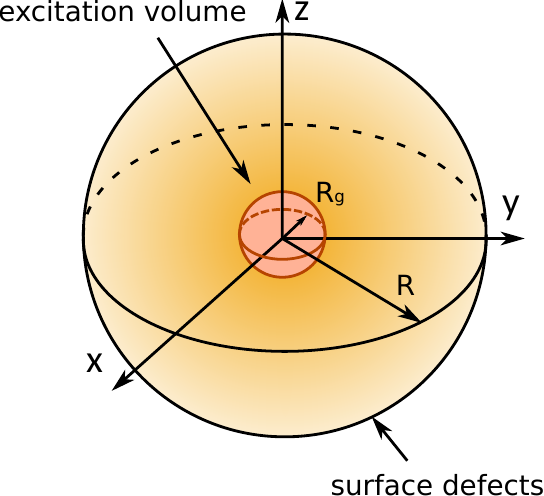}
    \caption{\label{fig:sphere}Optical beam focused at the center of a single
    grain of radius $R$ of a polycrystalline material, creating a spherical generation of
    carriers with radius $R_g$. The surface defects
    corresponds to the grain boundary.}
\end{figure}

In this section we consider a case more readily applicable to the analysis of a
single grain and its associated enclosing surface. The shape of the grains in a
polycrystalline sample strongly depends on the material and  the fabrication
process~\cite{Romeo_2001}. Here we limit ourselves to the
case of a grain with spherical symmetry of radius $R$, as depicted in
Fig.~\ref{fig:sphere}. This grain can be identified, for instance, in the bulk of a
sample after 3D mapping of the granular structure~\cite{Ludwig_2009}.
The geometry considered here also approximates systems for which the distance between the
sample surface and the buried sample-substrate interface is roughly the same as
the lateral grain size.  This would apply to CdTe photovoltaics, for which both
layer thickness and grain size are on the order of $1-3~\mathrm{\mu m}$. The surface
recombination velocity $S$ then represents an average over these
three distinct surfaces (sample surface, buried interface surface, and grain
boundary surface).  As in previous sections, we first solve the general diffusion
equation for this geometry, and then use the solution to propose procedures for
convenient data analysis.

\bigskip
The diffusion problem Eq.~(\ref{3Dproblem}) is now treated in spherical
coordinates with the boundary condition
\be
    D\frac{\partial n}{\partial r} (R,t) = -Sn(R,t).
    \label{BCsphere}
\ee
The generation volume is centered in the middle of the grain such that the
diffusion problem has radial symmetry. The solution of Eq.~(\ref{GF}) with the
boundary condition Eq.~(\ref{BCsphere}) is given by 
\be
    G(r,r',t) = \frac{e^{-t/\tau}}{2\pi R} \sum_{n>0} c_n
                \frac{\sin(\alpha_nr')}{r'}
                \frac{\sin(\alpha_nr)}{r}
                e^{-\alpha_n^2Dt},
    \label{G_final3}
\ee
with
\be
    c_n = \frac{\alpha_n^2R^2+(SR/D-1)^2}{\alpha_n^2R^2+(SR/D-1)SR/D}
\ee
where the coefficients $\alpha_n$ satisfy the transcendental relation
\be
    \tan(\alpha R) = \frac{\alpha R}{1-SR/D}.
    \label{trans2}
\ee
The details of the calculations are in Appendix~\ref{appendix3}. The system
being bounded in all directions, the time decay of the PL signal is now only
given by exponential terms. An approximation of the first non-zero root of
Eq.~(\ref{trans2}) was found~\cite{proceedings}
\be
    \tau_1 \approx \frac{R}{3S} + \frac{R^2}{\pi^2D}.
    \label{approx}
\ee
For collection times longer than $R^2/\pi^2D$, the Fourier expansion
in Eq.~(\ref{G_final3}) is dominated by its first term so that the PL signal
decays following a single exponential with an effective characteristic time
given by Eq.~(\ref{tau_PL}). 

We next propose a method for conveniently estimating $S$ and $D$, assuming that
the bulk lifetime is already known, by varying the generation/collection region
radius.  For a uniform spherical carrier generation/collection volume of radius
$R_g$ and a carrier generation per unit volume $g_0$. Integrating
Eq.~(\ref{G_final3}) twice over the sphere of radius $R_g$ yields the PL
intensity 
\be
    I(t) = \frac{g_0}{\tau_r}e^{-t/\tau}\frac{8\pi}{R}\sum_{n>0} \frac{c_n}{\alpha_n^4}
    A_n(R_g) e^{-\alpha_n^2Dt},
    \label{PL}
\ee
with
\be
    A_n(R_g) = \left[\sin(\alpha_nR_g) - \alpha_nR_g \cos(\alpha_nR_g)\right]^2.
    \label{An}
\ee
In the long time limit ($t\gg R^2/\pi^2D$) Eq.~(\ref{PL}) reads
\be
    I(t) = \frac{g_0}{\tau_r}\frac{8\pi}{R} 
    \frac{c_1}{\alpha_1^4} A_1(R_g) e^{-(1/\tau + \alpha_1^2D)t}.
    \label{I1}
\ee
We notice that the time constant of the signal decay is independent of the
excitation radius $R_g$.  Indeed, $R_g$ enters only in the prefactor $A_1(Rg)$.
For this reason, the ratio of PL signals from two different excitation radii is
a time-independent constant
\be
    \frac{I_1(t)/g_{0,1}}{I_2(t)/g_{0,2}} =
    \frac{A_1(R_{g,1})}{A_1(R_{g,2})},
    \label{ratio}
\ee
where $I_1(t)$ and $I_2(t)$ are two PL signals corresponding to
generation/collection volumes with radii $R_{g,1}$ and $R_{g,2}$, and
generations per unit volume $g_{0,1}$ and $g_{0,2}$.  We next outline how
Eq.~(\ref{ratio}) can be used to conveniently estimate $D$ and $S$.
\begin{figure}[t]
    \includegraphics[width=0.46\textwidth]{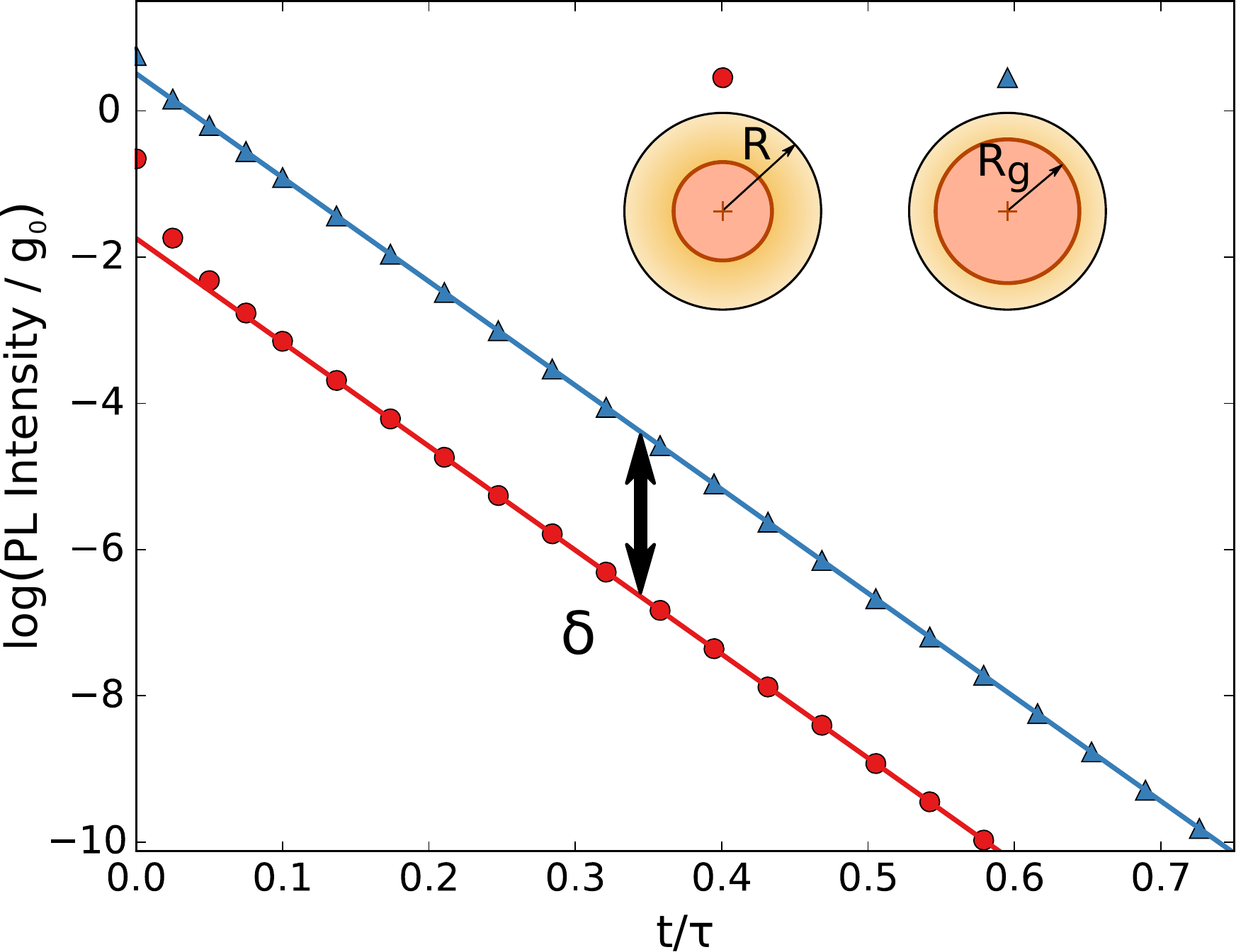}
    \caption{\label{sphere_scaling} PL intensities (normalized by the carrier
    generation $g_0$) as a function of time (in units of $\tau$) for a uniform
    spherical excitation. Symbols and lines correspond to Eq.~(\ref{PL})
    computed up to $n=200$ and Eq.~(\ref{I1}) respectively, with $R_g=0.5R$
    (dots, red), $R_g=0.8R$ (triangles, blue). The grain radius is
    $R=1~\mathrm{\mu m}$. We used $D=10~\mathrm{cm^2~s^{-1}},
    \tau=2~\mathrm{ns}$ and $S=5\times 10^{5}~\mathrm{cm~s^{-1}}$. Inset:
    schematic of the increase of the spot
    size (red) inside a grain (orange).}
\end{figure}
In Fig.~\ref{sphere_scaling}, we show the time-dependent PL intensities for two
excitations with different radii.  The symbols correspond to PL signals given by
the exact solution of Eq.~(\ref{PL}) with two different radii. The solid lines,
corresponding to the estimate Eq.~(\ref{I1}), show that the slopes of both
signals are the same, equal to $1/\tau+\alpha_1^2 D$. If the bulk lifetime is
already know, these slopes can be used to determine $\alpha_1^2 D$. We now need
to find $\alpha_1$.  We take the logarithm of Eq.~(\ref{ratio}) and replace
$A_1$ with Eq.~(\ref{An}) to obtain
\be
    \delta = 2\log\left[\frac{\sin(\alpha_1R_{g,1}) - \alpha_1R_{g,1}
    \cos(\alpha_1R_{g,1})}{\sin(\alpha_1R_{g,2}) - \alpha_1R_{g,2}
    \cos(\alpha_1R_{g,2})}\right],
    \label{delta}
\ee
where $\alpha_1$ is given by the approximation Eq.~(\ref{approx})
\be
    \alpha_1 = \frac{1}{\sqrt{D\left(\frac{R}{3S} + \frac{R^2}{\pi^2D}\right)}}.
    \label{a1}
\ee
Measuring the distance between both PL signals after the initial rapid decay,
gives $\delta$, as shown in Fig.~\ref{sphere_scaling}. Given $\delta$,
Eq.~(\ref{delta}) can be used to determine $\alpha_1$. Having determined
$\alpha_1^2D$ and $\alpha_1$, finding $D$ is trivial and  Eq.~(\ref{a1}) yields
the surface recombination
\be
    S = \frac{R}{3} \frac{\alpha_1^2 D}{1 - \left(\frac{\alpha_1 R}{\pi}
    \right)^2}.
\ee

\section{Conclusions}
The present work investigates the diffusion problem related to the two-photon
TRPL measurement technique.  This non-invasive and non-destructive optical
technique can
spatially resolve surface and subsurface features.  We have modeled the
transport of excess minority-carriers in three dimensions in the low injection
limit. A variety of boundary conditions were considered corresponding to
different experimental and material geometries. The results obtained here
suggested different routes for measuring physical parameters (diffusion
constant, bulk lifetime, surface recombination velocity) by making use of the PL
amplitude and time decay. Using the calculations done for a single planar
boundary, we fitted experimental data collected in the bulk and at the
substrate/active layer interface of heteroepitaxial CdTe/ZnTe/Si. We found that
the optics (spot size and location) significantly  impacts the precision of the
extracted physical parameters.  Finally, we have shown that PL signals are in
general {\it not purely} exponential. Purely exponential decays are recovered,
for instance, when the laser spot size is much greater than the diffusion
length. This amounts to assuming that all radiatively emitted photons are
collected.

\begin{acknowledgements}
The authors thank D.~Kuciauskas for providing the experimental data used in
Sec.~\ref{semi-infinite}. B.~G. acknowledges support under the Cooperative Research
Agreement between the University of Maryland and the National Institute of
Standards and Technology Center for Nanoscale Science and Technology, Award
70NANB10H193, through the University of Maryland.
\end{acknowledgements}

\begin{appendices}
\section{Derivation of Eq.(\ref{G_final})}
\label{appendix1}
We present here the steps followed to derive Eq.~(\ref{G_final}). Our starting
point is Eq.~(\ref{GF}) in Cartesian coordinates. Performing a Laplace transform
with respect to the time variable $t$ and two
Fourier transforms with respect to the space variables $x$ and $y$ leads to
\be
    \frac{\partial^2 G}{\partial z^2} - \mu^2 G = \frac{e^{-i(k_xx'+k_yy')}}{D}\delta(z-z')
\ee
with $\mu = \sqrt{k_x^2+k_y^2 + s/D + 1/(D\tau)}$. The general solution to this
equation is given by
\begin{align}
    &\forall z<z',\ \ G(k_x,x', k_y,y',z,z',s) = Ae^{\mu z} + Be^{-\mu z}\\
    &\forall z>z',\ \ G(k,z,s) = Ce^{-\mu z}
\end{align}
The continuity of $G$ at $z=z'$, the discontinuity of $\partial_z G$ at $z=z'$
and the boundary condition Eq.~(\ref{BC1}) fix the three constants $A, B, C$ so
that we obtain
\begin{align}
    G(k_x,x',k_y,y'&,z,z',s) = \frac{e^{-i(k_xx'+k_yy')}}{2D\mu}
    \nonumber\\
    &\times \left[e^{-\mu |z-z'|} + \frac{\mu D -
    S}{\mu D + S}e^{-\mu(z+z')} \right].
\end{align}
We now use the shift theorem of the Laplace transform,
\be
    \mathcal{L}\left[e^{-at}f(t)\right] = F(s+a),
\ee
to separate the Fourier and Laplace transforms
\begin{align}
    &G(k_x,x',k_y,y',z,z',t) = \frac{1}{2} e^{-(D(k_x^2+k_y^2) + 1/\tau)t - ik_xx' - ik_yy'}\nonumber\\
    &\times \int_{c-i\infty}^{c+i\infty}ds
    \frac{e^{sDt}}{\sqrt{s}}\left[e^{-|z-z'|\sqrt{s}} + \frac{D\sqrt{s} -
    S}{S\sqrt{s} + S}e^{-(z+z')\sqrt{s}}\right]
\end{align}
where we changed the variable $s/D$ for $s$. The inverse Fourier transform is
now straightforward and so is the inverse Laplace transform with the two
identities
\begin{align}
    &\mathcal{L}^{-1}\left[\frac{e^{-\sqrt{as}}}{\sqrt{s}} \right] =
    \frac{e^{-a/(4t)}}{\sqrt{\pi t}}\\
    &\mathcal{L}^{-1}\left[
    \frac{e^{-a\sqrt{s}}}{\sqrt{s}(\sqrt{s}+\beta)}\right] = e^{a\beta +
    \beta^2t} \mathrm{erfc}\left(\frac{a}{2\sqrt{t}} + \beta\sqrt{t}\right)
\end{align}
where $\mathrm{erfc}$ is the complementary error function. These transformations
yield Eq.~(\ref{G_final}).

\section{Derivation of Eq.(\ref{G_final2})}
\label{appendix2}
We now turn to the derivation of the Green's function Eq.~(\ref{G_final2}). Our
starting point is Eq.~(\ref{GF}) in Cartesian coordinates. Because the diffusion
remains in free space in the x- and y-direction and based on the previous
calculation, we introduce the following ansatz for the Green's function
\be
    G(x,x',y,y',z,z',t) =
    U(z,z',t)e^{-t/\tau}\frac{e^{-\frac{(x-x')^2}{4Dt}}}{2\sqrt{\pi Dt}}
    \frac{e^{-\frac{(y-y')^2}{4Dt}}}{2\sqrt{\pi Dt}}.
    \label{ansatz}
\ee
Substitution of Eq.~(\ref{ansatz}) into Eq.~(\ref{GF}) yields the
standard diffusion equation for $t>0$
\be
    D\frac{\partial^2 U}{\partial z^2} = \frac{\partial U}{\partial t}
    \label{diffusion}
\ee
with the initial profile $U(z,z',0)=\delta(z-z')$. The diffusion equation has been
extensively studied for various boundary conditions by Carslaw and Jaeger for
the heat flow problem~\cite{Carslaw_Jaeger}. The following form is assumed with
constants $A$ and $B$ determined by boundary conditions
\be
    U(z,z',t) = (A\cos(\alpha z) + B\sin(\alpha z))e^{-D\alpha^2t}.
\ee
The boundary conditions Eq.~(\ref{BC1}) and Eq.~(\ref{BC3}) yield
\begin{align}
    B = AS / (\alpha D)\\
    \tan(\alpha d) = \frac{\alpha (S+S')D}{\alpha^2D^2 - SS'}
    \label{Apptrans}
\end{align}
The general solution to Eq.~(\ref{diffusion}) is given by the summation over all
the roots of Eq.~(\ref{Apptrans})
\be
    U(z,z',t) = \sum_{n>0} c_n(z') (D\alpha_n \cos(\alpha_nz) + S
    \sin(\alpha_nz))e^{-\alpha_n^2Dt},
    \label{U}
\ee
where the coefficients $c_n(z')$ are given by
\be
    c_n(z') = \frac{\int_0^d \mathrm{d}z\ \delta(z-z') (D\alpha_n \cos(\alpha_nz) + S
    \sin(\alpha_nz))}{\int_0^d dz\ (D\alpha_n \cos(\alpha_nz) + S
    \sin(\alpha_nz))^2},
\ee
which gives after some algebra
\be
c_n(z') = \frac{2(D\alpha_n \cos(\alpha_nz') + S \sin(\alpha_nz'))}
{\frac{D^2\alpha_n^2+S^2}{D^2\alpha_n^2 + S'^2} \left[d(D^2\alpha_n^2+S'^2)+DS' \right] +
DS}.
\ee
The coefficient $A_n(z)$ given in Eq.~(\ref{series}) comes from a rewriting of
Eq.~(\ref{U}) and reads
\begin{align}
A_n(z) = \frac{\sqrt{2}(D\alpha_n \cos(\alpha_nz) + S \sin(\alpha_nz))}
{\sqrt{\frac{D^2\alpha_n^2+S^2}{D^2\alpha_n^2 + S'^2}\left[d(D^2\alpha_n^2+S'^2)+DS' \right] +
DS}}.
\end{align}

\section{Derivation of Eq.~(\ref{G_final3})}
\label{appendix3}
The derivation of Eq.~(\ref{G_final3}) follows the same procedure used in
App.~\ref{appendix2}, except that our starting point is Eq.~(\ref{GF}) in
spherical coordinates
\begin{align}
    -&\frac{\partial G}{\partial t}(r,r',t) + D \frac{1}{r^2}\frac{\partial}{\partial
    r}\left(r^2\frac{\partial G}{\partial r}\right)(r,r',t) - \frac{G(r,r',t)}{\tau} 
\nonumber\\ 
    &= \frac{\delta(r-r')}{4\pi r^2}\delta(t).
\end{align}
The Green's function takes the form
\be
    G(r,r',t) = \frac{U(r,r',t)}{r}e^{-t/\tau},
\ee
where $U$ satisfies the one-dimensional diffusion equation
\be
    D\frac{\partial^2 U}{\partial r^2} = \frac{\partial U}{\partial t}
\ee
with the boundary condition
\be
    D\frac{\partial U}{\partial r}(R,t) = \left(\frac{D}{R}-S\right)U(R,t)
\ee
and the initial profile
\be
    U(r,r',0) = \frac{\delta(r-r')}{4\pi r^2}.
    \label{eqU}
\ee
The details of solution of Eq.~(\ref{eqU}) can be found in Carslaw and Jaerger's
work~\cite{Carslaw_Jaeger}, and is given by the Fourier series
\be
    U(r,r',t) = \frac{1}{2\pi R} \sum_{n>0} A_n(r') \sin(\alpha_n r) e^{-\alpha_n^2Dt}
\ee
where
\be
    A_n(r') = \frac{D^2\alpha_n^2R^2+(SR-D)^2}{D^2\alpha_n^2R^2+SR(SR-D)}
          \frac{\sin(\alpha_n r')}{r'},
\ee
and the coefficients $\alpha_n$ are given by the transcendental relation
Eq.~(\ref{trans2}).

\end{appendices}

\bibliographystyle{my_apsrev}
\bibliography{references}

\end{document}